\documentclass{emulateapj}
\usepackage{apjfonts}

\newcommand{\msun}{M_\odot}

\newcommand{\gapprox}{\mathrel{\mathpalette\@versim>}}
\newcommand{\lapprox}{\mathrel{\mathpalette\@versim<}}
\newcommand{\propapprox}{\mathrel{\mathpalette\@versim\propto}}

\shorttitle{Kepler's SNR: Evidence for CSM Interaction}
\shortauthors{WILLIAMS ET AL.}

\begin{document}

\title{Dust in a Type Ia Supernova Progenitor: Spitzer Spectroscopy of
  Kepler's Supernova Remnant}

\author{Brian J. Williams,\altaffilmark{1,2,3}
Kazimierz J. Borkowski,\altaffilmark{1}
Stephen P. Reynolds,\altaffilmark{1}
Parviz Ghavamian,\altaffilmark{4}
William P. Blair,\altaffilmark{5}
Knox S. Long,\altaffilmark{6}
Ravi Sankrit\altaffilmark{7}
}

\altaffiltext{1}{Physics Dept., North Carolina State University,
    Raleigh, NC 27695-8202}
\altaffiltext{2}{NASA Goddard Space Flight Center, Greenbelt, MD, 20771; brian.j.williams@nasa.gov}
\altaffiltext{3}{Oak Ridge Associated Universities (ORAU), Oak Ridge, TN, 37831}
\altaffiltext{4}{Dept. of Physics, Astronomy, and Geosciences, Towson University, Towson, MD 21252}
\altaffiltext{5}{Dept. of Physics and Astronomy, Johns Hopkins University, 
    3400 N. Charles St., Baltimore, MD 21218-2686}
\altaffiltext{6}{STScI, 3700 San Martin Dr., Baltimore, MD 21218}
\altaffiltext{7}{SOFIA/USRA}

\begin{abstract}

Characterization of the relatively poorly-understood progenitor
systems of Type Ia supernovae is of great importance in astrophysics,
particularly given the important cosmological role that these
supernovae play. Kepler's Supernova Remnant, the result of a Type Ia
supernova, shows evidence for an interaction with a dense
circumstellar medium (CSM), suggesting a single-degenerate progenitor
system. We present 7.5-38 $\mu$m infrared (IR) spectra of the remnant,
obtained with the {\it Spitzer Space Telescope}, dominated by emission
from warm dust. Broad spectral features at 10 and 18 $\mu$m,
consistent with various silicate particles, are seen throughout. These
silicates were likely formed in the stellar outflow from the
progenitor system during the AGB stage of evolution, and imply an
oxygen-rich chemistry. In addition to silicate dust, a second
component, possibly carbonaceous dust, is necessary to account for the
short-wavelength IRS and IRAC data. This could imply a mixed chemistry
in the atmosphere of the progenitor system. However, non-spherical
metallic iron inclusions within silicate grains provide an alternative
solution. Models of collisionally-heated dust emission from fast
shocks ($>$ 1000 km s$^{-1}$) propagating into the CSM can reproduce
the majority of the emission associated with non-radiative filaments,
where dust temperatures are $\sim 80-100$ K, but fail to account for
the highest temperatures detected, in excess of 150 K. We find that
slower shocks (a few hundred km s$^{-1}$) into moderate density
material ($n_{0} \sim 50-250$ cm$^{-3}$) are the only viable source of
heating for this hottest dust. We confirm the finding of an overall
density gradient, with densities in the north being an order of
magnitude greater than those in the south.


\end{abstract}

\section{Introduction}
\label{intro}

Kepler's supernova remnant (SNR) is the remains of the supernova (SN)
of 1604 A.D., the third youngest known Galactic SNR, after G1.9+0.3
and Cas A. There is general consensus that Kepler resulted from a Type
Ia SN, albeit one with an unusually dense circumstellar medium (CSM)
(Blair et al. 2007, and references therein; hereafter B07), evidenced
by the extreme densities observed around Kepler ($\sim 2-100$
cm$^{-3}$), which, at $\gtrsim 500$ pc out of the galactic plane,
should have surrounding densities of $< 10^{-2}$ cm$^{-3}$
\citep{mckee77}, as well as the nitrogen overabundance seen in optical
spectra \citep{blair91}, and high ($\sim 200$ km s$^{-1}$)
blueshift. \citet{reynolds07} have suggested that Kepler may be a
member of an emerging class of ``prompt'' Type Ia SNe, evolving from
the main-sequence to SN in $< 5 \times 10^{8}$ yr. The distance to
Kepler was determined by \citet{sankrit05} to be 3.9 kpc, although
other estimates range from 3.3 kpc \citep{katsuda08} to $> 6$ kpc
\citep{chiotellis12}.

\begin{figure}
\includegraphics[width=17cm]{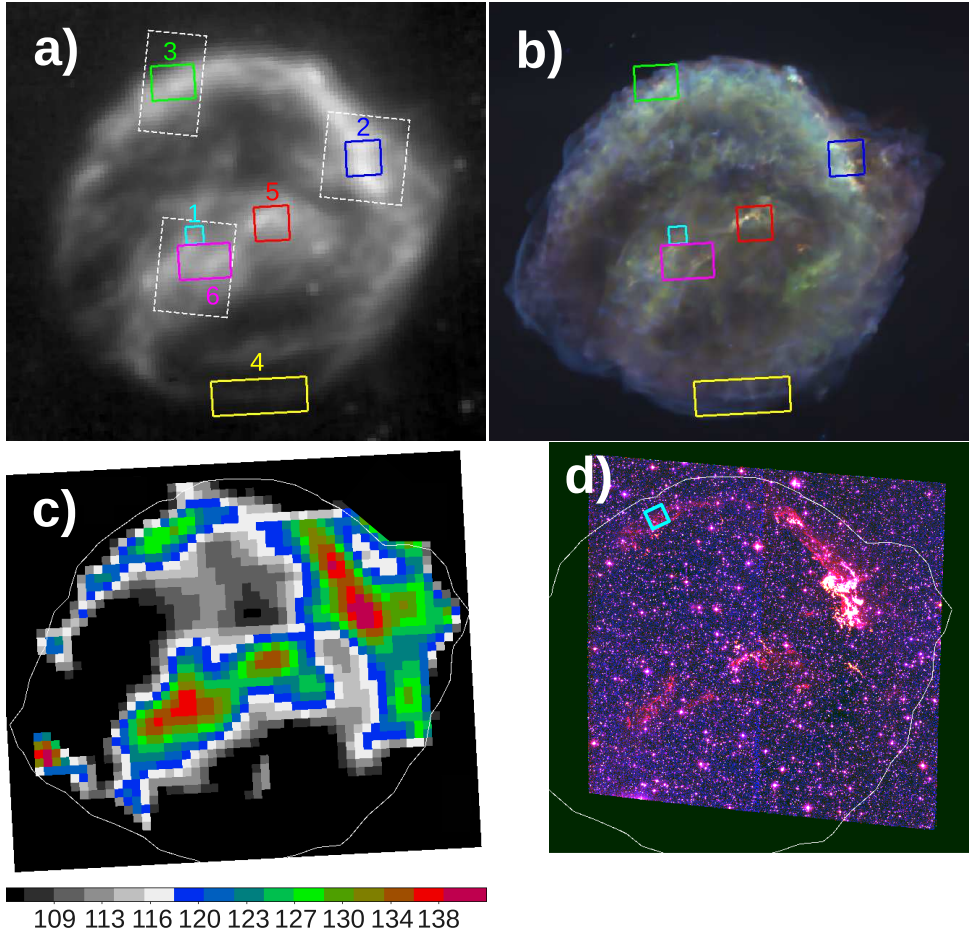}
\caption{a) {\it Spitzer} 24 $\mu$m image of Kepler (B07), overlaid
  with regions corresponding to spectra shown in Figure 2. Dashed
  white boxes represent locations of IRS SL maps. b) Three-color {\it
    Chandra} X-ray image \citep{reynolds07} with identical spectral
  regions overlaid. c) Temperature map (in K), as described in text,
  with scale indicated by color bar at the bottom. d) HST/ACS image
  with H$\alpha$ (red), [N II] (green), and [O III] (blue). Regions
  appearing white are bright in all bands, indicating radiative
  shocks. Cyan box shows extraction region for spectrum in Figure 5,
  discussed in section 4.  An outermost contour from just outside the
  forward shock is shown in panels c and d. N is up and E to the left
  in all images. Kepler is $\sim 3.8'$ in diameter.
\label{4panel}
}
\end{figure}

There is much debate in the general supernova literature about the
progenitor systems of Type Ia SNe, particularly as to whether they
result from single-degenerate systems (the explosion of a white dwarf
that has accreted matter close to the Chandrasekhar limit of $\sim 1.4
\msun$ from a companion) or double-degenerate (the merger of two
sub-Chandrasekhar mass white dwarfs). H$\alpha$ emission from
non-radiative shocks is seen at the periphery of the remnant, as we
show in Figure~\ref{4panel}. In the case of Kepler, the high densities
and presence of hydrogen in the CSM favor a single-degenerate
scenario, since a double white dwarf system, even if it could eject a
dense CSM, should not eject one rich in H. Furthermore, the presence
of dust in the CSM implies that the companion star was an AGB star,
since a main-sequence companion should not produce any significant
quantity of dust. \citet{chiotellis12} suggested that Kepler resulted
from a white dwarf explosion in a symbiotic binary, and is expanding
into the dense AGB wind from the progenitor system.

As the only known Type Ia SNR encountering the wind from its pre-SN
system, Kepler offers a unique opportunity to study the mass-loss from
evolved progenitor stars. AGB stars are typically classified as O-rich
or C-rich, with O-rich stars producing silicate dust and C-rich stars
producing carbonaceous dust in their outflows \citep{guha11}. Through
the interaction of Kepler's forward shock with this material, we can
constrain the type of star present in the progenitor system. The dust
in Kepler is known to be dominated by silicate grains
\citep{douvion01}, implying formation in an O-rich AGB star. 

The formation of dust in O-rich stars is still not entirely
understood; here is one promising scenario. Large, transparent
silicate grains form close to the stellar photosphere, within two
stellar radii. These are magnesium rich silicates, without iron
\citep{norris12}. As these grains are driven outward by the stellar
radiation pressure, temperatures fall and metallic iron condenses onto
the grains, forming ``dirty silicates'' with substantial opacity at
optical and near-IR wavelengths \citep{kemper02}. This scenario has
been confirmed in the laboratory by the identification of
circumstellar grains among GEMS (glass with embedded metal and
sulfides) grains; 1-6\% of GEMS grains are of circumstellar origin
\citep{keller11}. However, the question then becomes one of how the
presence of a white-dwarf companion can modify this process, both
before and after the supernova.

Prior to the explosion, accreting white dwarfs produce significant
amounts of energetic radiation and shocks, quite possibly enough to
dissociate molecules along the dust condensation sequence
\citep{bujarrabal10}. In an O-rich star, all C is locked up into CO,
but this dissociation (either via radiation or shocks) can also affect
CO molecules, leading to a mixed C/O chemistry in the outer dust
formation zone. This has been observed in planetary nebulae
\citep{guzman11}, and carbonaceous grains have also been detected
amongst silicate-rich dust in the symbiotic binary system CH Cyg
\citep{schild99}.

After the SN, the forward shock from the explosion will race through
the surrounding medium at several thousand km s$^{-1}$, heating the
gas to temperatures of $>10^{7}$ K. This hot plasma will heat grains
present in the CSM, destroying them via sputtering in the process
\citep{dwek92}.

In B07, we reported {\it Spitzer} imaging of Kepler, confirming the
presence of dense material ahead of the shock along with the existence
of a density gradient in the north-south direction, and placed an
upper limit of $< 0.1 \msun$ of dust present. Here, we report followup
observations with {\it Spitzer's} Infrared Spectrograph (IRS), to
study the processes listed above in the context of an SNR, and show
how the data can be interpreted in the context of the wind from a
pre-SN AGB star. Since Kepler's SNR represents the only known
opportunity to study dust in the CSM surrounding a Type Ia SN, any
characterization of the dust properties could have an impact on the
light echoes observed from Type Ia SNe \citep{patat06}.

\section{Observations}

We observed Kepler's SNR on 2009 March 29 with both orders of the
long-wavelength (14-38 $\mu$m), low-resolution (LL) IRS module and
order 1 (7.5-14 $\mu$m) of the short-wavelength, low-resolution (SL)
instrument. Both have a wavelength-dependent spectral resolving power
of $\lambda / \Delta\lambda = 64-128$. The entire remnant was mapped
with the LL instrument, while selected regions were mapped with the SL
module. The LL map consisted of four 30s cycles per position, with a
step size in the perpendicular direction of $9''$ after each 120s
observation. A total of 29 pointings in the perpendicular direction
were necessary. We then shifted the slits in the parallel direction by
$100''$ and repeated the process to cover the entire remnant with
sufficient background.

The three SL maps consisted of observations of varying depth to avoid
saturation. The center-east knots were observed with 22 pointings,
each with five cycles of 60s duration, yielding a total map size of
$56'' \times 43''$. The north rim position was covered with 17
pointings (seven 60s cycles each), for a total map size of $56''
\times 33''$. The northwest position, by far Kepler's brightest
region, was observed in 25 pointings (five 14s cycles each), for a
total map size of $56'' \times 49''$. A step size in the perpendicular
direction of $1.85''$ was used. The outline of the SL maps is shown on
Figure 1a.

The LL map allows creation of a three-dimensional (two spatial and one
wavelength) data cube using the IRS contributed software {\it
  CUBISM}\footnotemark\ \citep{smith07}, which allows spectral
extraction from any rectangular region. The pixel size of the LL IRS
instrument is $5.1''$, but {\it Spitzer's} point-spread function is $>
5.1''$ at wavelengths $> 20\ \mu$m. Thus, we require a minimum region
size of a 2x2 pixel square (10.2$''$x10.2$''$) when extracting LL
spectra, with the additional requirement that any 2x2 pixel square
falls within one IRS slit position, i.e. the two pixels do not fall on
different slit pointings. For the background, we use an average of
four off-source slit pointings (two north of the remnant, two
south). The mid-IR background surrounding Kepler is highly uniform.

\footnotetext[6]{http://irsa.ipac.caltech.edu/data/SPITZER/docs/dataanalysistools/tools/cubism/}

Extracting a full spectrum from 14 to 38 $\mu$m requires matching of
orders 1 \& 2. The flux levels of these orders generally did not match
perfectly, despite being extracted from identical regions. However,
since we are only interested in the shape of the spectra (see
Section~\ref{results}), we adjust the levels of the two orders to
match, using the order overlap of 19.5-21.3 $\mu$m. The relative
offset, a product of the inherent calibration uncertainties for each
order of IRS\footnotemark, varied from 5\% to 30\%, depending on the
region.

\footnotetext[7]{http://irsa.ipac.caltech.edu/data/SPITZER/docs/irs/features/}

The short wavelength end of LL order 1 (22-25 $\mu$m) contains a
consistent noise pattern of regularly-spaced spikes, present in every
spectrum. We used every {\it Spitzer} IRS data reduction program
available, but all efforts to remove these fringes were
unsuccessful. However, they have no discernible effect on the models
fits discussed in Section~\ref{results}, and we leave them in the data
shown.

\section{Results and Discussion}
\label{results}

\begin{figure}
\includegraphics[width=17cm]{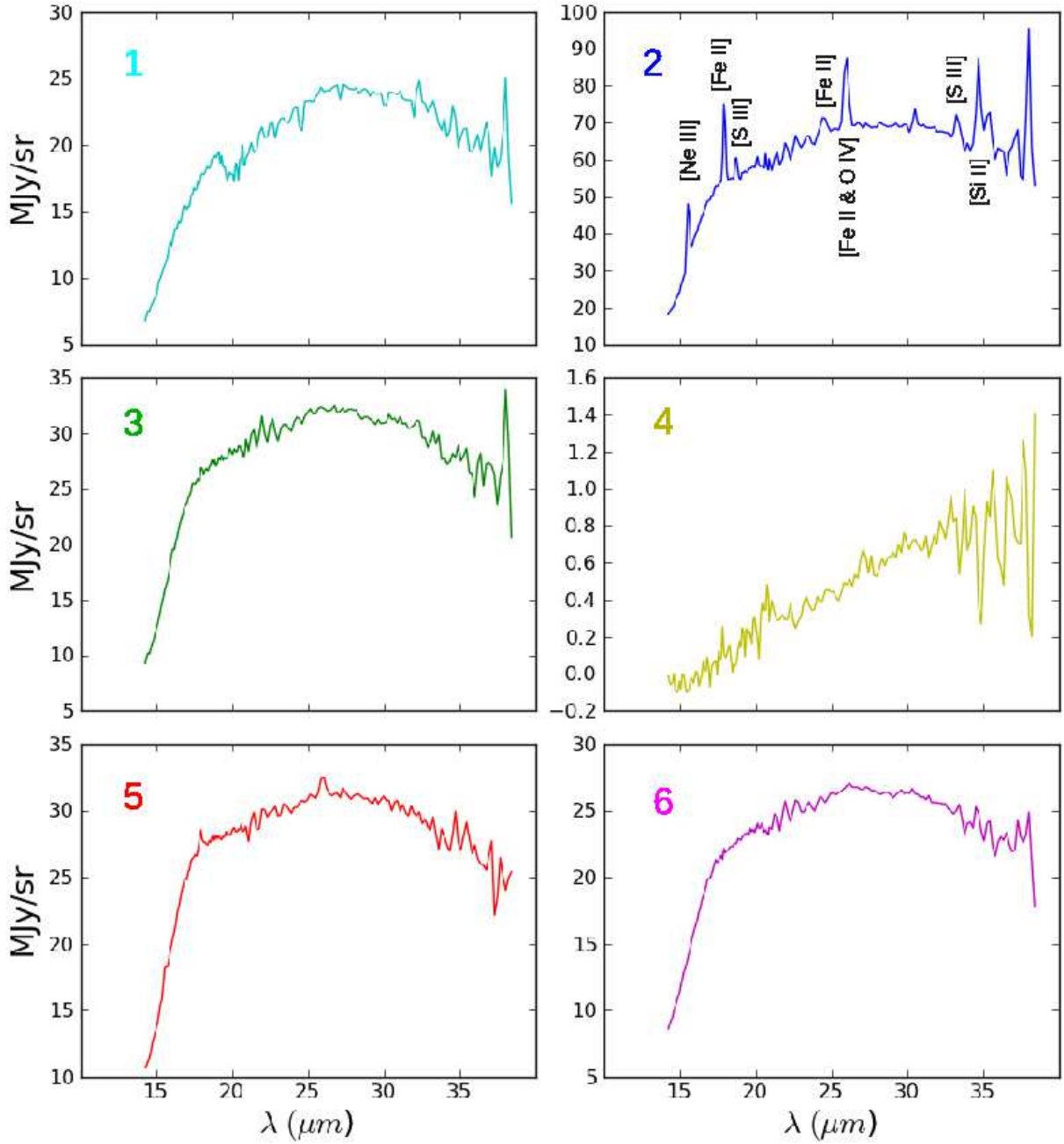}
\caption{IRS long-wavelength spectra of selected regions in Kepler,
  chosen to highlight the variation in dust throughout the
  remnant. Spectra are color-coded to regions shown in Figure 1 and
  plotted on a linear scale in surface brightness, averaged over the
  region.
\label{6plot}
}
\end{figure}

The spectral variation from place-to-place in Kepler is
significant. In Figure~\ref{4panel}, we show both the 24 $\mu$m {\it
  Multi-band Imaging Photometer for Spitzer} (MIPS) image of Kepler
and the {\it Chandra} X-ray image, overlaid with six different
spectral extraction regions, the spectra from which are shown in
Figure~\ref{6plot}. The exact locations of these six regions are given
in Table~\ref{regions}. Along the north and south rims (regions 3 and
4), spectra show a dust-dominated continuum with no lines, while in
the NW (region 2), lines from low-ionization state ions are present,
the only place in the remnant where emission lines are prominent. This
corresponds to the brightest region of radiative shocks seen in the
optical \citep{blair91}. Modeling of collisionally-heated dust grains
in the postshock environment provides strong constraints on the gas
density \citep{borkowski06,williams06}, but aside from a brief
discussion in Section~\ref{gradient}, we defer detailed spatial
mapping of densities to a future publication.

\subsection{Dust Temperature}

Determining dust temperatures in SNRs is complex, since different
grain sizes will be heated differently, with the smallest grains
undergoing extreme temperature fluctuations on short
timescales. However, we can make some simplifying assumptions and
create a proxy representative of temperature, in the same way that
``color temperatures'' are defined for stars. We created two
narrowband images of Kepler in spectral regions away from lines or
strong dust features ($28\ \mu$m and $16.2\ \mu$m), and divided these
images to make a ratio map. The value of the ratio at each image pixel
was fit to a model of silicate dust at a single temperature, assuming
a grain size of 0.05 $\mu$m. The output is shown in Figure 1c. Grain
``temperatures'' ranged from $<100$ K to $>150$ K, a high temperature
for SNR dust.

We made simple estimates of heating by UV photons from the radiative
shocks propagating into modest CSM densities. Specifically, we
consider the case of a shock of 500 km s$^{-1}$ encountering a medium
of $n_{0} = 250$ cm$^{-3}$, and assume that 100\% of the kinetic
energy of the shock is converted into radiation (this clearly provides
an upper limit, as such an approximation is unphysical). The radiation
energy density from this shock would be $\sim 1.2 \times 10^{-9}$ ergs
cm$^{-3}$. Using the formulae provided in \citet{draine11}, a
radiation field of this magnitude will heat typical ISM grains of
radius 0.1 $\mu$m to only $\sim 50$ K, far too low to account for dust
seen in Kepler. Collisional heating from hot gas in shocks is far more
efficient in most SNRs \citep{draine79}. There is a range of shock
speeds in Kepler, from the $\lesssim 100$ km s$^{-1}$ radiative shocks
in the NW to the $>3000$ km s$^{-1}$ non-radiative shocks in the south
rim. The latter, encountering the lowest density material ($n_{0}
\sim$ 1-5 cm$^{-3}$, B07) are capable of heating dust to 75-100
K. However, if we assume pressure equilibrium, i.e., $\rho v^{2}$ is
constant, then slower shocks of a several hundred km s$^{-1}$ should
be encountering material with densities of 50 to several $\times 100$
cm$^{-3}$. We use the grain heating models described in
\citet{borkowski06} and \citet{williams11} to examine the effects of
collisional heating on grains in these slower shocks. Briefly, grains
are heated via collisions with hot ions and electrons in the
post-shock gas, with final grain temperatures being a function of both
the plasma temperature and density. Shocks such as these can
collisionally heat grains to $>150$ K.

\subsection{North-South Density Gradient}
\label{gradient}

Kepler's north limb is significantly brighter than the south across
the EM spectrum. In B07, we attributed this to the northern shock
encountering material 4-9 times denser than that in the
south. \citet{katsuda08} measured the proper motions of X-ray emitting
filaments around the periphery, finding that shock velocities in the
northern filaments were 1.5-3 times slower than in the south, implying
a density ratio of 2-9.

We extracted IR spectra from near the locations where proper motions
were measured. Spectra of collisionally-heated dust are most sensitive
to the post-shock gas density, and by making reasonable assumptions
about the ion and electron temperature, we can fit a spectral model
where density is the only free parameter. We assume pure silicate dust
and use a $\chi^{2}$ minimization algorithm to fit the spectra from 21
to 33 $\mu$m, which we have found \citep{williams11} to be the most
reliable region of the spectrum to fit to obtain a gas density. In the
non-radiative sections of the north rim (region 3), we use the values
derived in B07 of $T_{i} = 8.9$ keV and $T_{e} = 1.4$ keV, obtaining a
post-shock density of $n_{H} = 42\ (41,43.5)$ cm$^{-3}$, with a
reduced $\chi^{2}$ of 1.3 (values in parentheses are the 90\%
confidence limits, but represent only the statistical errors on the
fit). In this model, $\sim 60\%$ of the dust is destroyed via
sputtering, roughly consistent with our initial modeling of the MIPS
data in B07.

\begin{figure}
\includegraphics[width=17cm]{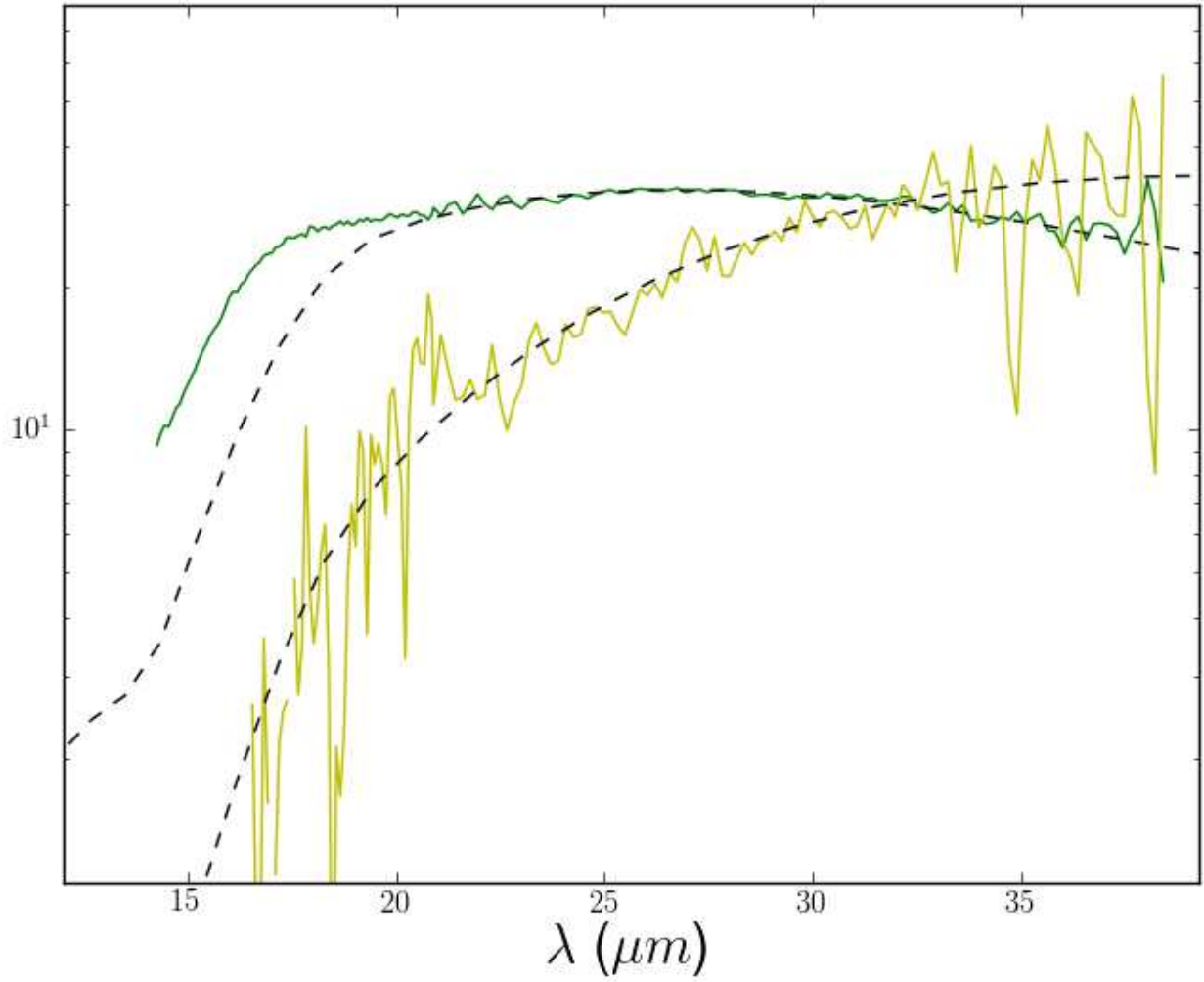}
\caption{LL ($\sim$ 14-38 $\mu$m) spectra from regions 3 (green) and 4
  (yellow), overlaid with dust model fits as described in the
  text. Region 4 is multiplied by a factor of 40 to be shown on the
  same scale. The fits assume pure silicate dust and are made only to
  the 21-33 $\mu$m region of the spectrum. From the fit to region 3,
  it can also be seen that the 18 $\mu$m silicate feature cannot be
  adequately reproduced by these models, which assume the optical
  constants of ``astronomical silicate'' from \citet{draine84}.
\label{northsouthfits}
}
\end{figure}

In the southernmost rim (region 4), X-rays are dominated by nonthermal
emission, and determining the plasma temperature is difficult. Using
the same temperatures as above, we find $n_{H} = 4$
cm$^{-3}$. However, it is likely, given the high shock speeds and low
densities, that far less temperature equilibration has taken place
between the ions and electrons \citep{ghavamian07}. If we thus assume
an ion temperature of 50 keV, allowing for heating of electrons
through Coulomb collisions alone, we derive a density of $n_{H} =
2.0\ (1.8,2.2)$ cm$^{-3}$ (reduced $\chi^{2}/$ = 0.99). Only 10\% of
the dust is destroyed via sputtering in this region, owing to the much
lower gas densities. We show the fits from both regions, 3 \& 4, in
Figure~\ref{northsouthfits}. Slightly farther south, in the region
that in X-rays is dominated by thin, synchrotron emitting filaments
\citep{reynolds07}, we find a weak dust continuum that is best fit
with a density of 1 cm$^{-3}$. At {\it Spitzer's} resolution, it is
difficult to determine if this continuum is real or simply due to the
wings of the point-spread function from the brighter emission just
above. Thus, we consider $n_{H} = 1$ cm$^{-3}$ to be an upper limit
for the southernmost rim. The addition of graphite grains to these
models (see Section~\ref{short}) increases the derived densities by
$\sim 20$\%, but the overall density gradient is unchanged.

\subsection{Silicate Dust}

\begin{figure}
\includegraphics[width=17cm]{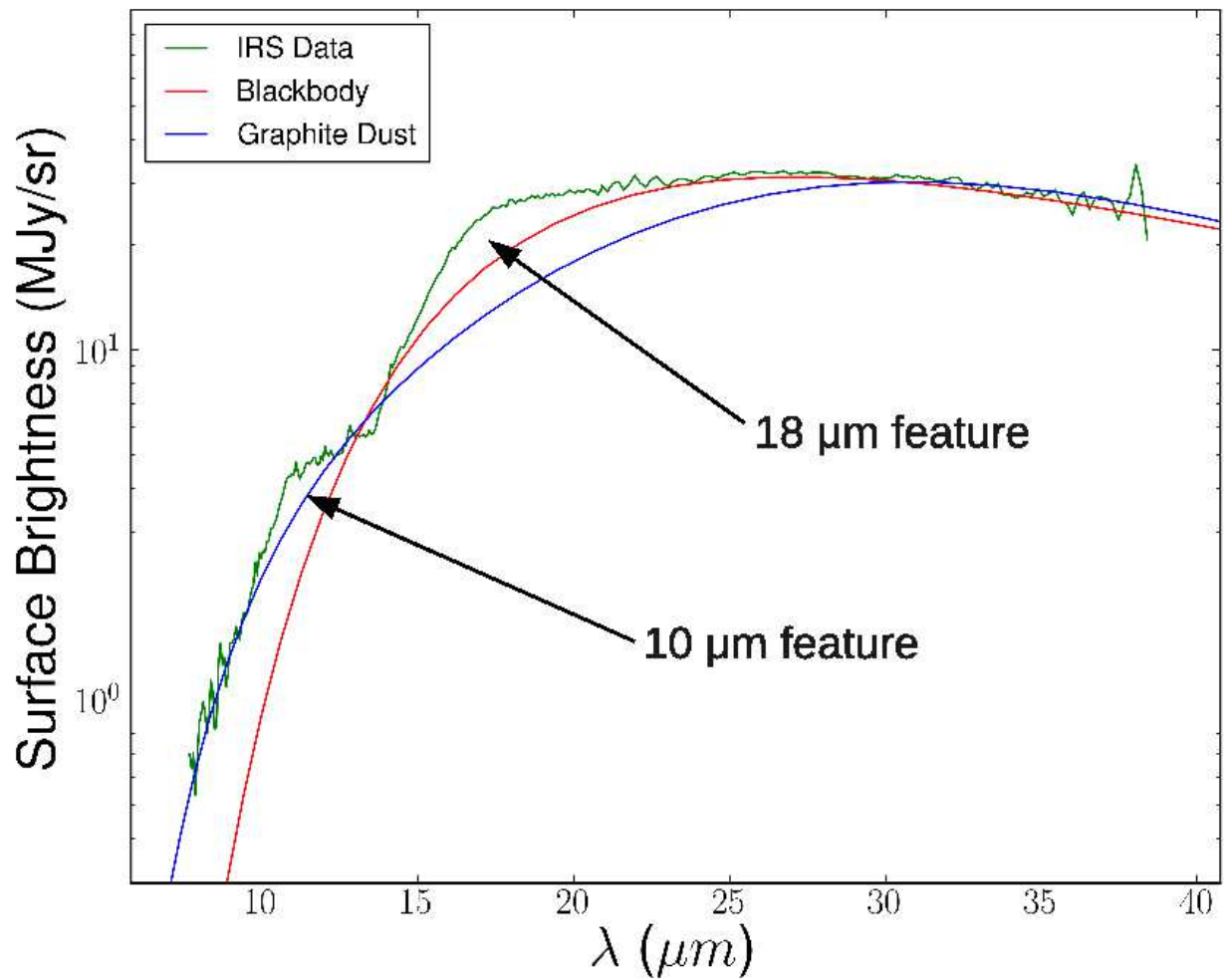}
\caption{The SL \& LL (7.5-40 $\mu$m) spectrum of region 3, plotted on
  a logarithmic scale. A single-temperature blackbody model and a
  single-temperature 0.05 $\mu$m graphite grain model are fit to the
  data and overlaid to show the existence of the broad 10 and 18
  $\mu$m silicate emission features.
\label{totalbb}
}
\end{figure}

Strong features, most easily attributable to silicate dust, exist in
the spectra shown in Figures 2-5 at $\sim 10$ and 18 $\mu$m. In
Figure~\ref{totalbb}, we show the total 7.5-40 $\mu$m spectrum from
region 3, overlaid with two different model fits to the data: a simple
blackbody model (T $\sim 120$ K) and an emission model for graphite
grains. The broad emission features, visible in the data at $\sim 10$
$\mu$m and $\sim 18 \mu$m are easily seen in excess of either of these
two models. While these silicate features are often observed in
absorption in cold dust of the ISM, they are routinely observed in
emission in places where dust is warm, such as SNRs
\citep{rho08,dwek10} or the optically-thin dust shells of O-rich AGB
stars with moderate mass-loss rates \citep{guha11}. \citet{douvion01}
first identified these features in Kepler in {\it Infrared Space
  Observatory} spectra, reporting that they could be fit with standard
silicate dust with optical constants from \citet{draine84}. We find no
evidence for silicate absorption features anywhere in the remnant.

On the smallest measurable scales ($10.2'' \times 10.2''$), a peculiar
feature is present in the spectra of about two dozen regions within
the remnant. In these regions, the 18 $\mu$m silicate feature appears
to be so strong that it peaks above the level of the continuum,
dipping back down sharply on the long wavelength side. An example of a
spectrum like this is shown in the upper left panel of
Figure~\ref{6plot}. We have no complete explanation for this
feature. When slightly larger spatial extraction regions are used, the
feature disappears. We examined the individual spectral observations
that went into making the cube using SPICE, the Spitzer Custom
Extraction Software, to search for evidence of the feature's
existence, and found that the appearance and disappearance of the
feature depended entirely on the spectral extraction method used. The
effect is strongest in regions of the remnant where strong gradients
in the IR surface brightness exist. Because of this, we believe it is
most likely that it is an artifact of the spectral extraction
algorithms used by {\it CUBISM} on the smallest spatial scales, and
not an absorption feature or abnormally strong unknown emission
feature.

\subsubsection{Possible Ejecta Dust?}

Two possibilities exist for the dust seen in Kepler: newly formed
ejecta dust or dust created in the AGB wind of the progenitor
system. Ejecta dust is typically associated with core-collapse SNe,
but recent theoretical work by \citet{nozawa11} has shown that it may
be possible for up to 0.2 $\msun$ of dust to form in the ejecta of
Type Ia SNe. However, to date, there have been {\em no} detections of
ejecta dust firmly associated with Type Ia SNe. \citet{gomez12} report
a small amount ($\sim 3 \times 10^{-3}\ \msun$) of dust detected in
{\it Herschel} observations (comparable to what we found in B07), and
mention the possibility that they cannot rule out, based on their {\it
  Herschel} data alone, a contribution from ejecta dust, but they do
not claim detection.

Since the first detection of ejecta dust in a Type Ia SNR would be of
extreme interest, we have searched extensively for any signs of it in
Kepler. We have gone about this in two ways. First, \citet{burkey12}
separate the X-ray data from {\it Chandra} \citep{reynolds07} into
regions dominated by ejecta emission, thermal CSM emission, and
nonthermal synchrotron radiation. A detailed morphological comparison
between these data and ours shows no obvious evidence for a
correlation between IR emission and X-ray ejecta dominated
regions. While we cannot rule out an ejecta dust contribution to the
IR spectra (complications from projection effects cannot be ignored),
it does not appear to be significant.

Secondly, another comparison can be made between the location of dust
emission and the emission line maps produced by CUBISM, a technique
used for a spectroscopic study of Cas A by \citet{rho08}, where they
use [Fe II] and other emission lines. Kepler is $\sim 20$ times
fainter than Cas A and the maps produced are of much lower
signal-to-noise, and the [Fe II] lines, which can come from either
ejecta or shocked or photo-ionized gas are the only ones strong enough
to make a spectral map. The [Fe II] emission line maps at 17.9 and 26
$\mu$m show clear emission from several regions in Kepler, but again,
no obvious correlation exists between the regions of strong line
emission and peculiar dust spectral features, as is the case in Cas
A. Again, while we cannot rule out an ejecta dust component to the IR
spectra, we unfortunately have no direct evidence for it.

Lastly, as we point out in B07, there is a clear morphological
similarity between dust emission seen with all of {\it Spitzer's}
instruments (the {\it Infrared Array Camera}; IRAC, MIPS, and IRS) and
the H$\alpha$ emission shown in Figure~\ref{4panel}. Since this
H$\alpha$ emission is associated with the non-radiative shocks at the
shock front, we conclude that the dust is associated with the swept-up
CSM.

\subsubsection{CSM Dust}

It thus appears that the shock is primarily encountering dust formed
in the outflow from the progenitor system, either the precursor star
itself or the donor. AGB stars produce a slow, dense wind that is
ideal for the formation of grains. The thermodynamic condensation
sequence for dust formation in the atmosphere of an O-rich AGB star is
complex and not fully understood (see Section~\ref{intro}), but the
end result in high mass-loss rate outflows is a significant amount of
amorphous silicate dust that we also see in Kepler \citep{waters11}.

While it is clear that projection effects ensure that the spectra in
Kepler sample regions of differing temperature, density, and
composition, we can nonetheless arrive at some general conclusions
even without detailed modeling. First, the spectra from the central
knots, shown in Figure~\ref{6plot}, are remarkably similar to the
spectrum from the north rim, consistent with the hypothesis that
emission from these central knots results from CSM dust. Second,
although the overall shape of the continuum at long wavelengths is
relatively well-fit, the 18 $\mu$m silicate feature is {\em not}
well-fit with the astronomical silicate model of \citet{draine84},
which places the centroid of the feature at $\sim 18.2\ \mu$m. The
silicate features seen in the spectra of Kepler in the north,
northwest, and central knots are centered at $\sim 17$ $\mu$m. This is
most easily seen in Figure~\ref{northsouthfits}, where a fit to the
long-wavelength spectrum of region 3 significantly underpredicts
emission from 15-18 $\mu$m. To further test this conclusion, we follow
the technique of \citet{guha11} of fitting a blackbody model to the
underlying continuum (using a least-squares routine) and dividing this
out of the spectrum. What remains is the absorption efficiency of the
silicate dust grain as a function of wavelength. While the 10 $\mu$m
feature is in the correct place (to the extent that our
short-wavelength data will allow us to determine this), the 18 $\mu$m
feature is not, and is much closer to 17 $\mu$m. We find {\em no} dust
temperature at which the optical constants of \citet{draine84} can
account for the features observed. This may point to a different
chemistry within the silicate grains. This is perhaps not entirely
surprising, as there are systematic differences between CSM and ISM
silicates features \citep{ossenkopf92}. Furthermore, a great variation
in silicate features is seen in AGB outflows \citep{henning10},
suggesting marked differences in physical and chemical properties
among CSM grains. We also note that a similar mismatch of the 18
$\mu$m feature was observed in the central region of M81
\citep{smith10}, where the feature peaks at 17.2 $\mu$m. The authors
there found a better (but far from perfect) match when using the
dielectric function of amorphous olivine as measured in a laboratory
by \citet{dorschner95}. But distinguishing among various types of
silicates in Kepler is a difficult task that requires separate efforts
involving more sophisticated spectral modeling and reliance on more
extensive laboratory measurements (see \citet{speck11} for recent
experimental studies of amorphous silicates). 

We find no indication of the strong spectral features produced by
crystalline silicate particles, such as those seen in the spectra
around cool, evolved O-rich stars \citep{morris08,henning10}. These
features are generally most prominent in high mass-loss rate ($\ga
10^{-5} M_ \odot$ yr$^{-1}$) AGB outflows, so it is possible that the
mass-loss rate of the progenitor system of Kepler was not extreme. But
crystalline silicates may be common as well at low mass-loss rates,
just harder to detect (e.g., see de Vries et al.~2010 and references
therein), and the signal-to-noise ratio in Kepler may make any such
detection impossible. Crystalline silicates are rarely detected in SNR
(for a counter-example where they are detected, see \citet{koo11} and
compare the spectra from MSH 15-52 to that from Kepler), and a likely
explanation for this is that impacts with thermal ions amorphize
crystalline silicates, changing their chemical structure. For the fast
non-radiative shock in the north of Kepler, with $T_i = 8.9$ keV and
$n_H \sim 40$ cm$^{-3}$, the time required for the complete
amorphization is only several years for grains $\sim 0.1$ $\mu$m in
radius, based on critical proton and $\alpha$ particle doses in
\citet{glauser09}. In the south, densities are an order of magnitude
lower than in the north, and the amorphization timescale is inversely
proportional to density. Crystalline silicates might survive under
these conditions, but again, would be harder to detect because of low
IR fluxes. Crystalline material might also survive in cores of large
grains, shielded from energetic ions by overlying amorphized surface
layers, particularly in slow shocks driven into the dense CSM where
ions do not have enough energy to penetrate into grain
cores. \citet{glauser09} showed that such partially crystalline grains
can still produce detectable crystalline emission features, but we see
no evidence for them in Kepler.

\section{The Near-IR Excess}
\label{short}

\begin{figure}
\includegraphics[width=15cm]{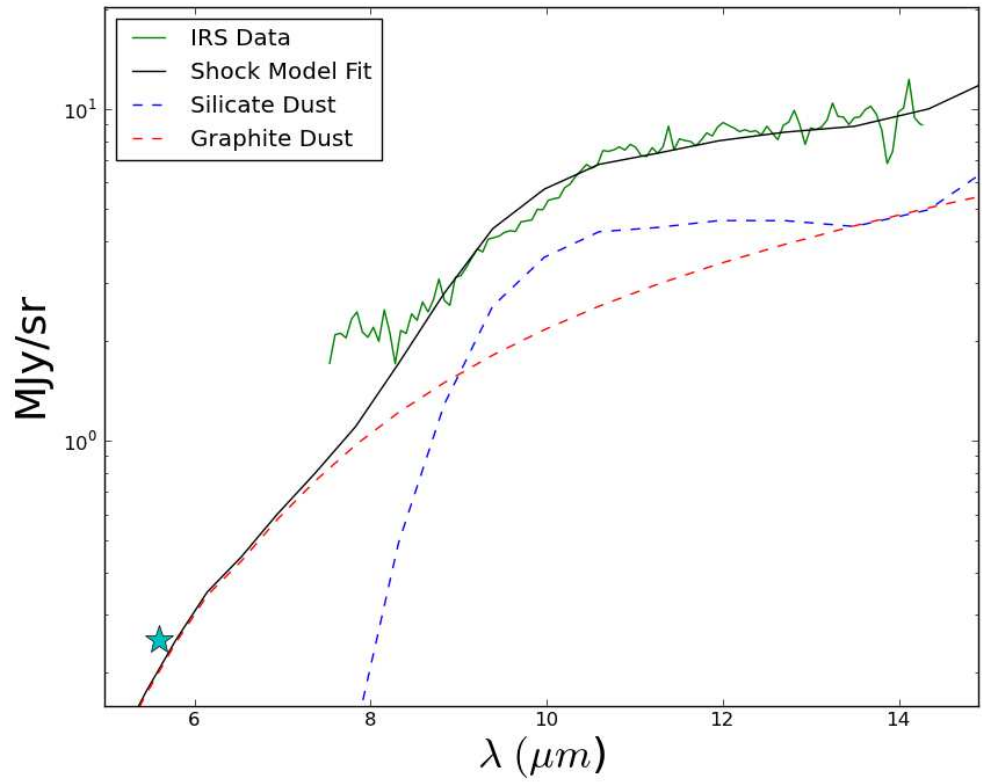}
\caption{SL (7.5-14 $\mu$m) spectrum (in surface brightness, on
  logarithmic scale) of NE emission knot (cyan region in Figure 1d),
  overlaid with shock model fit as described in text. The IRAC 5.6
  $\mu$m data point, used as a constraint to the fit, is shown as a
  star.
\label{si_gr}
}
\end{figure}

Since dust in Kepler exhibits prominent silicate emission features, it
must have condensed in a stellar outflow with dominant O-rich
chemistry (C/O ratio $\la 1$). However, perhaps the most intriguing
spectral feature of Kepler is that while the spectra are clearly
dominated by silicate grains, these alone {\em cannot} account for the
short-wavelength data, including the IRAC fluxes. We use IRAC data
obtained as part of our earlier survey of Kepler with both IRAC and
MIPS (PI W. Blair, program ID 3413, see B07). We show this in
Figure~\ref{si_gr}, the IRS short-low spectrum from the bright NE knot
(rectangular region in Figure 1d). This NE knot is not unique; other
locations in Kepler show a similar near-IR excess. A local background,
subtracted from both sides of the knot, was used to highlight the
hottest dust. For the IRAC data, we measured the flux from an
identical region ({\it CUBISM} allows creation of region files that
can be read by ds9) at 5.6 and 8.0 $\mu$m. The 8 $\mu$m flux was
roughly consistent with the IRS data, and thus does not provide an
additional constraint, but the 5.6 $\mu$m flux provides a strong
confirmation of the near-IR excess. At 3.6 and 4.5 $\mu$m, Kepler is
too faint and the knot is undetected.

A model fit to the data with pure silicates underpredicts the IRAC 5.6
$\mu$m emission by two orders of magnitude. This near-IR excess cannot
be explained by synchrotron emission, nor by an additional hot
silicate component. This near-IR excess has been a long-standing issue
in observations of both the ISM and envelopes around late-type stars
\citep{ossenkopf92}. An additional component of graphite dust provides
a better explanation than pure silicates, as shown in
Figure~\ref{si_gr} (though the fit is far from perfect at the shortest
wavelengths), which results from a shock model ($v_{s} = 500$ km
s$^{-1}$, $n_{H} = 250$ cm$^{-3}$) of collisionally-heated dust, where
silicates and graphites are assumed to have a power-law size
distribution ($\alpha = -3.5$) from 1 nm to 0.5 $\mu$m and a mass
ratio of silicates to graphites of two. Such a model, which still
underpredicts the observed emission, requires a mixed C/O chemistry in
the AGB progenitor system. This is possible in rare S-type AGB stars
where $C/O \approx 1$ \citep{ferrarotti02}. Mixed C/O chemistry can
operate in S stars if shocks are generated within their winds by
stellar pulsations, leading to formation of both carbonaceous and
silicate grains \citep{cherchneff06,smolders10}. If dust in Kepler was
produced by an S-type AGB star, this might explain the presence of
silicates together with carbonaceous grains. Unfortunately, we are
unable to determine the C/O ratio from UV and X-ray observations
because of the high interstellar absorption towards the remnant.
 
We also consider more common M-type ($C/O < 1$) AGB outflows, where in
single stars nearly all C is locked in CO molecules. In a
single-degenerate Type Ia progenitor system, an accreting white dwarf
produces energetic UV/X-ray radiation, and thermonuclear outbursts
drive shocks into the ambient CSM. A fraction of CO molecules can be
dissociated by shocks or energetic photons in the dust-forming regions
in these systems \citep{bujarrabal10}. This creates favorable
conditions for formation of carbonaceous dust even if $C/O < 1$. While
quantitative models of carbonaceous dust formation in O-rich symbiotic
systems are lacking, \citet{guzman11} demonstrate how energetic
photons can lead to mixed carbon-oxygen chemistry in O-rich outflows
of Galactic bulge planetary nebulae. Their models show that formation
of hydrocarbon molecules, precursors to PAHs and hydrogenated
carbonaceous dust, is more efficient at larger $C/O$ ratios.

However, the near-IR excess could also be produced by metallic iron
(or various iron oxides). A similar near-IR excess, with respect to
models of pure silicate dust, was found by \citet{kemper02} in their
study of the O-rich AGB star OH 127.8+0.0. They examined the chemical
abundances of various materials likely to be present in AGB
atmospheres, considering several potential (non-carbonaceous) dust
species that could form in addition to silicates, finding that
non-spherical metallic Fe inclusions within the silicate particles of
$\sim 4\%$ (by mass) could best account for the 3-8 $\mu$m
continuum. Such grains have recently been identified in terrestrial
laboratory samples by \citet{keller11}, who conclude, based on their
isotopic compositions, that some of these grains could originate from
AGB stars. By comparing the dust mass absorption coefficients
\citep{kemper02} for these ``dirty silicates'' to the near-IR excess
we see in Kepler, we find that a similar amount of non-spherical
Fe-inclusions ($\sim 5\%$ by mass) in amorphous silicate particles can
account for the IRAC 5.6 $\mu$m flux. This model does not require a
mixed C/O chemistry in the progenitor system. This is also true of
other possible models with separate populations of non-silicate
Fe-bearing grains (e.g., see models of \citet{dwek10} for SN 1987A
which shows a similar near-IR excess).

\section{Summary}

While other remnants of Type Ia SNe at comparable ages to Kepler are
interacting with ISM \citep{williams11}, Kepler is unique in
interacting with CSM, providing a laboratory to study the interaction
of the forward shock with the material ejected by the progenitor
system during its AGB phase \citep{chiotellis12}. While Kepler is
roughly spherical in shape, we find more than an order of magnitude
density contrast between the northern and southern rims. The strong
silicate features observed here are indicative of the wind from an
O-rich AGB star. However, there is clear evidence in the
short-wavelength emission for an additional dust component, with
either carbonaceous dust or metallic-Fe inclusions within amorphous
silicates providing a possible explanation. The presence of
carbonaceous grains would imply a mixed oxygen-carbon chemistry in the
progenitor system. There is no morphological evidence for newly-formed
ejecta dust. Small scale studies of the silicate features seen here,
the 9.7 $\mu$m feature in particular, will be an excellent
observational goal for future IR telescopes, such as the {\it James
  Webb Space Telescope}, which can search for crystalline silicates
and help distinguish between carbonaceous and iron grains.

\newpage

\begin{deluxetable}{lccc}
\vspace{-0.2truein}
\tablecolumns{4}
\tablewidth{0pc}
\tabletypesize{\footnotesize}
\tablecaption{Spatial Locations of Spectral Extractions}
\tablehead{
\colhead{Region} & R.A. (J2000) & Dec. (J2000) & Size (arcsec)}

\startdata
1 & 17:30:43.1 & -21:29:30.9 & 10.2 $\times$ 10.2\\
2 & 17:30:36.0 & -21:28:45.5 & 20.4 $\times$ 20.4\\
3 & 17:30:44.1 & -21:28:00.7 & 25.5 $\times$ 20.4\\
4 & 17:30:40.4 & -21:31:06.2 & 56 $\times$ 20.4\\
5 & 17:30:39.9 & -21:29:24.1 & 20.4 $\times$ 20.4\\
6 & 17:30:42.7 & -21:29:46.6 & 30.6 $\times$ 20.4\\

\enddata

\tablecomments{Positions listed are the center of the region. Sizes of
  regions are given as length $\times$ height.}
\label{regions}
\end{deluxetable}

\newpage
\clearpage

\begin{thebibliography}


\bibitem[Blair et al.(1991)]{blair91}
Blair, W.P., Long, K.S., \& Vancura, O. 1991, ApJ, 366, 484

\bibitem[Blair et al.(2007)]{blair07}
Blair, W.P., Ghavamian, P., Long, K.S., Williams, B.J., Borkowski, K.J., Reynolds, S.P., \& Sankrit, R. 2007, ApJ, 662, 998

\bibitem[Borkowski et al.(2006)]{borkowski06} 
Borkowski, K.J., et al. 2006, ApJ, 642, L141

\bibitem[Bujarrabal et al.(2010)]{bujarrabal10}
Bujarrabal, V., Miko\l ajewska, J., Alcolea, J., \& Quintana-Lacaci, G.
2010, A\&A, 516, A19

\bibitem[Burkey et al.(2012)]{burkey12}
Burkey, M.T., et al. 2012, in preparation

\bibitem[Cherchneff(2006)]{cherchneff06}
Cherchneff, I. 2006, A\&A, 456, 1001

\bibitem[Chiotellis et al.(2012)]{chiotellis12}
Chiotellis, A., Schure, K.M., \& Vink, J. 2011, A\&A, 537, 139

\bibitem[Draine(2011)]{draine11}
Draine, B.T. 2011, {\it Physics of the Intstellar and Intergalactic Medium}, Princeton Univ. Press



\bibitem[de Vries et al.(2010)]{devries10}
de Vries, B. L., Min, M., Waters, L. B. F. M., Blommaert,J. A. D. L., \& Kemper, F. 2010, A\&A, 516, A86

\bibitem[Dorschner et al.(1995)]{dorschner95}
Dorschner, J., Begemann, B., Henning, T., Jager, C., \& Mutschke, H. 1995, A \& A, 300, 503

\bibitem[Douvion et al.(2001)]{douvion01}
Douvion, T., Lagage, P.O., Cesarsky, C.J., \& Dwek, E. 2001, A\&A, 373, 281

\bibitem[Draine \& Salpeter(1979)]{draine79}
Draine, B.T., \& Salpeter, E.E. 1979, ApJ, 231, 438

\bibitem[Draine \& Lee(1984)]{draine84}
Draine, B.T., \& Lee, H.M. 1984, ApJ, 285, 89

\bibitem[Dwek \& Arendt(1992)]{dwek92}
Dwek, E. \& Arendt, R.G. 1992, ARA\&A, 30,11

\bibitem[Dwek et al.(2010)]{dwek10}
Dwek, E., et al. 2010, ApJ, 722, 425


\bibitem[Ferrarotti \& Gail(2002)]{ferrarotti02}
Ferrarotti, A. S., \& Gail, H.-P. 2002, A\&A, 382, 256

\bibitem[Ghavamian et al.(2007)]{ghavamian07}
Ghavamian, P., Laming, J.M., \& Rakowski, C.E. 2007, ApJ, 654, 69

\bibitem[Glauser et al.(2009)]{glauser09}
Glauser, A. M., et al. 2009, A\&A, 508, 247

\bibitem[Gomez et al.(2012)]{gomez12}
Gomez, H.L., et al. 2012, MNRAS, 420, 3557

\bibitem[Guha Niyogi et al.(2011)]{guha11}
Guha Niyogi, S., Speck, A.K., \& Onaka, T. 2011, ApJ, 733, 93

\bibitem[Guzm\'{a}n-Ram\'{i}rez et al.(2011)]{guzman11}
Guzm\'{a}n-Ram\'{i}rez, L., Zijlstra, A. A., N\'{i}chuim\'{i}n, R., 
Gesicki, K., Lagadec, E., Millar, T. J., \&  Woods, P. M.
2011, MNRAS, 414, 1667



\bibitem[Henning(2010)]{henning10}
Henning, T. 2010, ARA\&A, 48, 21

\bibitem[Katsuda et al.(2008)]{katsuda08} 
Katsuda, S., Tsunemi, H., Uchida, H., \& Kimura, M., ApJ, 689, 225

\bibitem[Keller \& Messenger(2011)]{keller11}
Keller, L.P. \& Messenger, S. 2011, GeCoA, 75, 5336

\bibitem[Kemper et al.(2002)]{kemper02}
Kemper, F., de Koter, A., Waters, L.B.F.M., Bouwman, J., \& Tielens, A.G.G.M. 2002, A\&A, 384, 585

\bibitem[Koo et al.(2011)]{koo11}
Koo, B.-C., et al. 2011, ApJ, 732, 6


\bibitem[McKee \& Ostriker(1977)]{mckee77}
McKee, C.F. \& Ostriker, J.P. 1977, ApJ, 218, 148

\bibitem[Morris(2008)]{morris08}
Morris, P.W. 2008, in: IAU Symposium, vol. 250, p. 361

\bibitem[Nozawa et al.(2011)]{nozawa11}
Nozawa, T., Maeda, K., Kozasa, T., Masaomi, T., Nomoto, K., \& Umeda, H. 2011, ApJ, 736, 45

\bibitem[Norris et al.(2012)]{norris12}
Norris, B.R.M., Tuthill, P.G., Ireland, M.J., Lacour, S., Zijlstra, A.A., Lykou, F., Evans, T.M., Stewart, P., \& Bedding, T.R. 2012, Nature, 484, 220

\bibitem[Ossenkopf et al.(1992)]{ossenkopf92}
Ossenkopf, V., Henning, Th., \& Mathis, J.S. 1992, A\&A, 261, 567

\bibitem[Patat et al.(2006)]{patat06}
Patat, F., Benetti, S., Cappellaro, E., \& Turatto, M. 2006, MNRAS, 369, 1949

\bibitem[Reynolds et al.(2007)]{reynolds07}
Reynolds, S.P., Borkowski, K.J., Hwang, U., Hughes, J.P., Badenes, C., Laming, J.M., \& Blondin, J.M. 2007, ApJ, 668, L135

\bibitem[Rho et al.(2008)]{rho08}
Rho, J., Kozasa, T., Reach, W.T., Smith, J.D., Rudnick, L., DeLaney, T., Ennis, J.A., Gomez, H., \& Tappe, A. 2008, ApJ 673, 271

\bibitem[Schild et al.(1999)]{schild99}
Schild, H., Dumm, T., Folini, D., Nussbaumer, H., \& Schmutz, W. 1999, ed. P. Cox \& M.F. Kessler, in The Universe as Seen by ISO, ESASP, 427, 397

\bibitem[Sankrit et al.(2005)]{sankrit05}
Sankrit, R., Blair, W.P., Delaney, T., Rudnick, L., Harrus, I.M., \& Ennis, J.A. 2005, AdSpR, 35 1027


\bibitem[Smith et al.(2007)]{smith07}
Smith, J.D.T., Armus, L., Dale, D.A., Roussel, H., Sheth, K., Buckalew, B.A., Helou, G., \& Kennicutt, R.C. 2007, PASP, 119, 1133

\bibitem[Smith et al.(2010)]{smith10}
Smith, H.A., et al. 2010, ApJ, 716, 490

\bibitem[Smolders et al.(2010)]{smolders10}
Smolders, K., et al. 2010, A\&A, 514, L1

\bibitem[Speck et al.(2011)]{speck11}
Speck, A.K., et al. 2011, ApJ, 740, 93


\bibitem[Waters(2011)]{waters11} Waters, L.B.F.M., in: Why Galaxies
  Care About AGB Stars II: Shining Examples and Common Inhabitants,
  eds. F. Kerschbaum, T. Lebzelter, \& R.F. Wing, ASP Conference
  Series, Vol. 445


\bibitem[Williams et al.(2006)]{williams06}
Williams, B.J., et al. 2006, ApJ, 652, L33

\bibitem[Williams et al.(2011)]{williams11}
Williams, B.J., et al. 2011, ApJ, 729, 65



\end{thebibliography}
\end{document}